\documentclass[pre,
twocolumn]{revtex4}
\begin{document}





\title{Comment on ``Observation of a Push Force on the End Face of a
Nanometer Silica Filament Exerted by Outgoing Light"}


\author{Iver Brevik}\email{iver.h.brevik@ntnu.no}
\affiliation{Department of Energy and Process Engineering,
Norwegian University of Science and Technology, N-7491 Trondheim,
Norway}


\maketitle

The Abraham-Minkowski electromagnetic energy-momentum tensor
problem has been on the agenda since about 1910. The recent
experiment of She {\it et al.} \cite{she08} is in this connection
of interest, as it shows how the radiation force from a
low-intensity laser yields an inward push force on the end face of
a vertical fiber.

But does this experiment measure electromagnetic momentum? In our
opinion the answer is no. What is
 detected is merely the  electromagnetic Abraham-Minkowski force
density ${\bf f}^{AM} = -(\varepsilon_0/2)E^2 {\bf \nabla} n^2$ in
the surface layer of the filament (or in other regions where $n$
varies). This is not related to the electromagnetic momentum in
itself.  The electromagnetic force density is ${\bf f}= {\bf
f}^{AM}+[(n^2-1)/c^2]\partial /\partial t({\bf E\times H)}$, and
electromagnetic momentum does not appear until the second term in
this expression. This is   the Abraham term. It is in principle
measurable  although
 it is usually small; moreover it simply fluctuates out when averaged
over an optical period in a stationary beam.

For illustration, let us assume that a short laser pulse with
energy $\cal H$ falls from vacuum towards the entrance surface of
a free-standing fiber ( we ignore gravity). If there is an
antireflection film of refractive index $\sqrt n$ on the surface,
$\cal H$ is the energy of the pulse in the medium also. The
impulse imparted to the surface because of the surface force
${f}^{AM}$ is $G_{surf}={\cal{H}} (n-1)/c$, directed against the
beam if $n>1$. When the pulse leaves at the exit surface, a
corresponding reverse impulse is imparted. In the Abraham case,
one has to take into account the mechanical momentum $G_{mech}^A$
caused by the Abraham term also. One finds $G_{mech}^A={\cal
H}(n^2-1)/nc$. The resulting longitudinal displacement of the
fiber because of the sum $G_{surf}+G_{mech}^A$ becomes $\Delta
x^A=({\cal H}/c^2\mu)(n-1)$, where $\mu=M/L$ is the mass per unit
length. As discussed on p. 189 in Ref.~\cite{brevik79}, $\Delta
x^A \sim~$1 pm or less, and is clearly non-observable.

In the present case, the fiber is fixed at the upper end. There
will be a downward directed impulse imparted to the fiber at the
lower end when the pulse leaves. It is very small: taking the flux
to be 10 mW and the pulse duration to be 270 ms, we get ${\cal{H}}
= 2.7$ mJ resulting in $G_{surf} = 4.5$ pN$\cdot$s if $n=1.5$.
Because of elasticity, there will be an upward directed recoil in
the fiber. (Sideways motion may result from non-axisymmetric
elastic  conditions.)

We propose  finally  a modification of the
 experiment that might  be capable of detecting the Abraham force after all (cf. also  p.~191
 of \cite{brevik79}): Let a long fiber of length $L$ be wound up on a
 drum of radius $R$ and of small weight.  Suspend the drum as a  vertical torsional
 pendulum such that it can oscillate about the $z$ axis with an
 eigenfrequency $\omega_0$. Then  send an {\it intensity  modulated}    optical
  wave through the fiber, such that its harmonic component has the same frequency
 $\omega_0$. We take the incident energy flux in vacuum to be
 $P^{(i)}=P_0\cos^2(\frac{1}{2}k_0x-\frac{1}{2}\omega_0t)$,
 where $k_0=\omega_0/c$, $x$ is the longitudinal coordinate,
 and $P_0$ is the unmodulated energy flux averaged over an
 optical period. With antireflection films on the end surfaces  we obtain in the Abraham case a longitudinal force
 $F^A=P_0[(n-1)/cn]\sin(\frac{1}{2}nk_0L)\sin(\frac{1}{2}nk_0L-\omega_0t)$,
whereas in the Minkowski case $F^M=-nF^A$. These forces give
 rise to measurable axial torques $N_z$ on the drum. Assuming the sheet
 of fiber on the drum to be thin we
 obtain, when setting $\sin(\frac{1}{2}nk_0L) \approx \frac{1}{2}nk_0L$, in
 the Abraham case $N_z^A=[(n-1)/2c^2]RLP_0\omega_0\sin(\frac{1}{2}nk_0L-\omega_0t)$.
In the Minkowski case, $N_z^M=-nN_z^A$. The two predictions are
 thus quite different.

For definiteness,assume  that a  YAG laser
  at $1.06~\mu$m  produces the incident beam. Assume
 that a high power of $P_0=1~$kW can be transmitted through the
 fiber, and neglect any losses. Then, with $L=100~$m, $n=1.5$, $R=10~$cm,
 $\omega_0=10~{\rm s}^{-1}$, we obtain for the predicted torque amplitudes
 $N_z^A=2.8\times 10^{-13}~$Nm, $N_z^M=4.2\times
 10^{-13}~$Nm.

 The above amplitudes are less than those of Ref.~\cite{roosen74}
($10^{-12}~$Nm), but of the same order of magnitude as in
Ref.~\cite{jones54}.
  Actually, they are greater than
those in the classic experiment of Ref.~\cite{beth36}
($10^{-16}~$Nm). Realization of our proposed experiment appears
difficult but not impossible.

Finally, it should be mentioned that our  Einstein-box argument
above implicitly assumed wide lateral dimensions for the pulse.
Cf. also the  Comment of Mansuripur on this point
\cite{mansuripur09}.

\end{document}